\documentclass[twocolumn,showpacs,preprintnumbers,amsmath,amssymb,prl]{revtex4}

\usepackage{dcolumn}
\usepackage{bm}

\usepackage{amssymb}
\usepackage{amsmath}
\usepackage[dvips]{graphicx}
\usepackage{color}
\usepackage{multirow}
\usepackage{epstopdf}
\usepackage{changes}
\usepackage{comment}

\usepackage{todonotes}

\begin{document}


\title{Statistical Mechanics Model for the Dynamics of Collective Epigenetic Histone Modification}

\author{Hang Zhang$^{1}$,  Xiao-Jun Tian$^1$,  Abhishek Mukhopadhyay$^2$, K.S. Kim$^3$,Jianhua Xing$^{1,2,4}$}
\email{jxing@vt.edu}
\affiliation{$^1$Department of Biological Sciences, Virginia Tech, Blacksburg, Virginia, 24061-0406, USA\\
$^2$Department of Physics, Virginia Tech, Blacksburg, Virginia, 24061-0406, USA\\
$^3$Lawrence Livermore National Laboratory and University of California, Livermore, California, 94550, USA \\
$^4$Beijing Computational Science Research Center, Beijing 100084, China
}

\begin{abstract}
\noindent
Epigenetic histone modifications play an important role in the maintenance of different cell phenotypes. The exact molecular mechanism for inheritance of the modification patterns over cell generations remains elusive. We construct a Potts-type model based on experimentally observed nearest-neighbor enzyme lateral interactions and nucleosome covalent modification state biased enzyme recruitment. The model can lead to effective nonlocal interactions among nucleosomes suggested in previous theoretical studies, and epigenetic memory is robustly inheritable against stochastic cellular processes.
\end{abstract}

\maketitle

Nucleosomes are basic organizational units of chromatin in eukaryotic cells. A typical nucleosome has approximately 147 base pairs wrapped around a histone octamer and are interconnected by linker DNA  of varying length (see Fig. 1) \cite{Alberts2008,Gaffney2012}. Covalent modifications of several amino acid residues on the histone core can lead to either active or repressive gene expression activities \cite{Bannister2011}. A dynamic equilibrium in the nucleosome modification state is attained due to a `tug-of-war' between the associated covalent mark addition and removal enzymes  \cite{Steffen2012}. The system may show a bistable behavior due to coexistence of repressive  and active epigenetic states for different copies of a gene  within the same cell \cite{Dodd2007}.

\noindent

Experiments suggest that at least some of the nucleosome covalent patterns can be transmitted over a number of generations \cite{Alberts2008}.  Although the actual mechanism for this epigenetic memory is unclear, a simple rule-based model by Dodd \textit{et al.} \cite{Dodd2007} shows  that robust bistability requires  cooperative effects beyond neighboring nucleosomes, which they suggest is due to compact chromatin structures. Subsequent theoretical studies  on yeast chromatin silencing \cite{Sedighi2007}, mouse stem cell differentiation \cite{binder2013transcriptional}, and plant flowering regulation \cite{Angel2011} also conclude that this nonlocal cooperativity is necessary for generating stable epigenetic memory.

\begin{figure}
    \includegraphics*[width=3.375in, keepaspectratio=true]{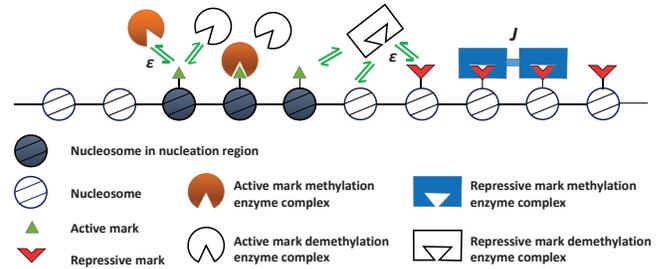}
     \caption{(Color online) Schematic illustration of the model. $\epsilon$ denotes enzyme binding energy, $J$ denotes enzyme lateral interaction energy}
     \label{fig:model}
\end{figure}

In recent years molecular details on nucleosome covalent modification dynamics have been extensively studied. Measurements show that the typical residence time of a modification enzyme on chromatin is within sub-seconds to a few minutes \cite{Steffen2012}. Experimental observations also suggest that a modified nucleosome may have higher binding affinity for the corresponding enzymes  \cite{Bannister2011, Berr2011, Schmitges2011, Kouzarides2007}.  Another interesting observation is that a nucleosome bound modification enzyme complex  laterally interacts  with another  bound to neighboring nucleosomes \cite{Schmitges2011, Ruthenburg2007,Canzio2011}. Although the functional consequences  of these  interactions on epigenetic dynamics are unclear, recent work suggests that increased enzyme lateral interactions lead to sustained repression or activation of genes, and  cancer cells show mutations linked with such lateral interactions  \cite{Norwood2006,so2003dimerization}.

In this work we construct a theoretical model aiming to bridge the gap between detailed molecular events occurring at the sub-second time scale,  and  the long-time scale epigenetic change dynamics that  is typically in days or longer. To be specific we focus on lysine 4 (active) and lysine 9 (repressive) methylation on histone H3, although we expect the mechanism discussed here can be general.

Consider a gene with $N$ nucleosomes, as shown in Fig. \ref{fig:model}. Each nucleosome can be in one of the three covalent states: repressively modified ($s =-1$), unmodified (0), and actively modified (1). Here for simplicity we only consider  one covalent modification site on each histone octamer, and do not distinguish multiple modification (\textit{e.g.}, mono-, di-, and tri-methylation)  states. Our model is flexible enough to admit straightforward extensions with increasing complexity.  Each nucleosome can be in one of the enzyme binding states with corresponding binding energies:  empty ($\sigma = 1, \epsilon = \epsilon_{1s} = 0$), repressive modification enzyme bound ($\sigma = 2, \epsilon = \epsilon_{2s}$), repressive modification removal enzyme bound ($\sigma = 3, \epsilon = \epsilon_{3s}$), active modification enzyme bound ($\sigma = 4, \epsilon = \epsilon_{4s}$), active modification removal enzyme bound ($\sigma = 5, \epsilon = \epsilon_{5s}$). To account for the $s$-dependence of binding affinity, we assume that the binding energies for the adding/removing enzymes to  a nucleosome bearing the corresponding (or antagonizing) mark are
$\Delta\epsilon$ lower (or higher) than those binding to an unmodified nucleosome. Furthermore if two neighboring (\textit{i}-{th} and (\textit{i+1})-{th})  nucleosomes are both bound, the binding enzymes interact laterally with energy $J_{\sigma_i\sigma_{i+1}}$, otherwise $J_{\sigma_i=1,\sigma_{i+1}} = J_{\sigma_i,\sigma_{i+1}=1}=0$.
The above background enzyme-nucleosome binding has no DNA sequence specificity, and the corresponding binding energies estimated from experimental data are weak. It is suggested that transcription factors and other molecules recruit the enzymes to bind on specific nucleosomes with significantly stronger binding affinity  \cite{buscaino2013distinct}. DNA sequence elements, e.g. CpG islands, have also been shown to have higher but less sequence-specific enzyme binding affinity \cite{Allen2006, Mendenhall2010, Ku2008, figueiredo2012hp1a, voigt2013double}. Therefore, we denote a special nucleation region of nucleosomes (for H3K4me3 and H3K9me3 centered around the transcription start site (TSS))  with lower binding energies. We will index the middle nucleosome within this region as 0,  those on its left negative, and those on its right positive. More details of the model can be found in the online supporting text. Specifically there is a nucleosome-free region near the TSS \cite{Zhang2011}, and some DNA distortion may be needed.

The overall  model has the structure of a coupled two-layer Potts model. Modification of the $s$-state requires the corresponding enzyme bound, and the enzyme binding energy is  $s$-dependent.  The $s$-state of a nucleosome can also be changed to 0 due to stochastic exchange with unmarked histones in solution with a rate of about once per 100 minutes  \cite{Deal2010}, and random replacement with 50\% probability by an unmarked histone every cell division \cite{Probst2009}, which is around 20 hours \cite{Kumei1989} for mammalian cells. The total number of states is $15^N$, which is computationally prohibitive for direct dynamic simulations. Given the clear time scale separation between the enzyme binding/unbinding events and other processes,  we treat the former as an equilibrium process, and simulate others stochastically, as described below.

\begin{table}
\caption{\label{Table 1} Model parameters}
\begin{ruledtabular}
\begin{tabular}{lc}
methylation free energy of binding within& -1 \\
nucleation region  $\epsilon_{\sigma 0}, \sigma=2,4$  &\\
demethylation enzyme free energy of binding within & 1 \\
nucleation region $\epsilon_{\sigma 0}, \sigma=3,5$  &\\
methylation  enzyme free energy of binding outside & 3\\
nucleation region $\epsilon_{\sigma 0}, \sigma=2,4$  & \\
demethylation enzyme free energy of binding outside& 3\\
nucleation region $\epsilon_{\sigma 0}, \sigma=3,5$  & \\
$s$ state related free energy of binding bias $\Delta\epsilon$ & 2 \\
\multirow{2}{7cm}{lateral interaction between two identical enzyme molecules $J_{\alpha\alpha}$} & \multirow{2}{*}{3}\\\\
\multirow{2}{7cm}{lateral interaction between two different enzyme molecules $J_{\alpha\beta}, \alpha\neq\beta$}& \multirow{2}{*}{-2}\\\\
enzymatic interaction rate $v_{\alpha\rightarrow\beta}$ & 1.5/hour\\
histone exchange rate $d$ & 0.6/hour\\
cell cycle time & 20 hours\\
\end{tabular}
\end{ruledtabular}
\end{table}

The interactions between covalent modification enzymes and a collection of nucleosomes at given $s$-states  can be described by the following Hamiltonian
\begin{eqnarray}
 H_s = \sum_{i=1}^N \epsilon_{\sigma_is_i} - \sum_{i=1}^{N-1} J_{\sigma_i\sigma_{i+1}}.
\end{eqnarray}
 Throughout this work energy is given in units of $k_BT$, where $k_B$ is the Boltzmann constant, and $T$ is the temperature.
 The partition function is,
$Z_s = \sum_{\{\sigma\}}  \exp(- H_s) = Tr[{\mathbf T}_1\cdots {\mathbf T}_N]$,
where the transfer matrix $\mathbf{T}_i$ has elements $(T_i)_{\alpha \beta}= \exp  (-\frac{1}{2} ( \epsilon_{\alpha s_i}+\epsilon_{\beta s_{i+1}}) + J_{\alpha\beta} )$ for $i=1,\cdots, N-1$,
and $(T_{N})_{\alpha\beta}= \exp (-\frac{1}{2}( \epsilon_{\alpha s_N}+\epsilon_{\beta s_1}  ))$. For notational simplicity, we omit the \textit{s}-dependence of the transfer matrices.
Then the probability of finding site $i$ in state $\sigma_i$ is
$P_{\sigma_i}(\{s\}) = Tr \left[  \mathbf{T}_1\cdots\mathbf{T}_{i-2}  \mathbf{G}_{\sigma_i} \mathbf{T}_{i+1}\cdots\mathbf{T}_{N} \right]/Z_s$,
\label{eqn:P_calculation}
where $(G_{\sigma_i})_{\alpha\beta} =(T_{i-1})_{\alpha, \sigma_i} (T_{i})_{\sigma_i, \beta} $ except  $(G_{\sigma_1})_{\alpha\beta} =(T_{N})_{\alpha,\sigma_1} (T_{1})_{\sigma_1,\beta} $.

Our overall simulation procedure is as follows: at each step, we first calculate $ \{P_{\sigma_i}\} (\{ s\})$ of each nucleosome, 
then update the $s$ state using the Gillespie algorithm with the possible events including enzymatic reaction on nucleosome $i$ with rate
$k_i = \delta_{s_i,0} (v_{0\rightarrow -1}P_{\sigma_i=2} (\{ s\})+v_{0\rightarrow 1}P_{\sigma_i=4} (\{ s\})) + \delta_{s_i,-1} v_{-1\rightarrow 0} P_{\sigma_i=3} (\{ s\}) + \delta_{s_i,1} v_{1\rightarrow 0}P_{\sigma_i=5} (\{ s\}) $, where $\delta$ is the Kronecker delta function, and histone exchange $(s_i\rightarrow 0)$  with rate $d$;  at every cell cycle (20h), the $s$-state of each nucleosome has 50\% probability to be reset to 0.

\begin{figure} 
    \centering\includegraphics*[width=3.3in, keepaspectratio=true]{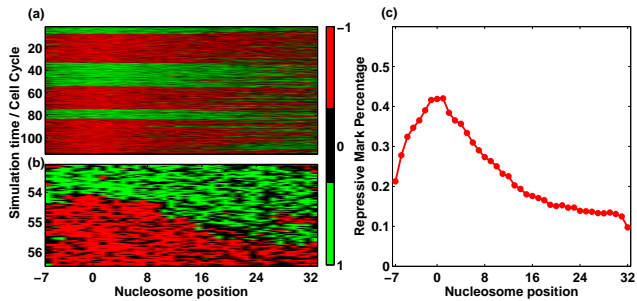} \caption{(Color online) Simulation results using model parameters corresponding to Oct4 (Table 1). (a): Heat map representation of a typical trajectory. (b): Zoom-in of the heat map in (a) showing epigenetic state transition. (c): Probability of observing repressive marks at different nucleosome sites. The nucleation region is at nucleosomes -1, 0, and 1.}
     \label{fig:oct4}
\end{figure}

We select the model parameters roughly representing the gene Oct4, one of the core genes maintaining \mbox{cell} pluripotent stemness \cite{Boyer2005}, and  the one monitored by Hathaway \textit{et al.} \cite{Hathaway2012}.  Figure \ref{fig:oct4}(a) shows a typical simulated trajectory. Despite large fluctuations, a block of nucleosomes centered around the nucleation region show collective dynamics, with occasional switches between repressive (light gray in print, green online) and active (dark gray in print, red online) states. Closer examination of a switching event (Fig. \ref{fig:oct4}(b)) shows that a cluster of nucleosomes with the same type of mark initially form around the nucleation region, then propagate steadily outwards. Indeed, the epigenetic state can be reversed by artificially changing the nucleosome marks within the nucleation region rather than outside this region, consistent with the experiment done by Hathaway et al. \cite{Hathaway2012}.

The (N=40) simulation is generated by the following steps: the values of free energy of binding are estimated from measured enzyme bound fraction and concentrations \cite{Steffen2012}, values of $J$ are chosen to reproduce the bell-like shaped histone methylation pattern centered around the nucleation region with a half-height width of about 10 nucleosomes (Fig. \ref{fig:oct4}(c)) \cite{Hathaway2012}, and $k$ is chosen to reproduce the observed $\sim$4 days of transition time from an active to a silent gene state \cite{Hathaway2012}. This set of model parameters, as summarized in Table 1, serves as the starting point for analyzing the model dependence on parameters. For simplicity, in this work we assume that the boundaries of the nucleosome region under study are occupied by insulating elements \cite{Zhao2004, Bushey2008} which could  prevent spreading of the epigenetic modifications beyond the region. The qualitative results of this work are not affected by using alternative periodic boundary conditions.

\begin{figure}
\includegraphics*[width=3.3in, keepaspectratio=true]{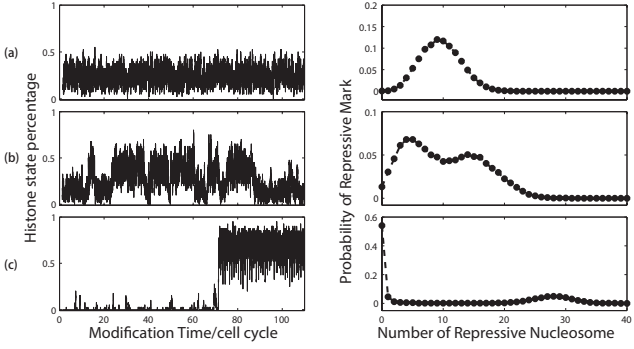}
 \caption{ Typical trajectories of the fraction of nucleosomes with repressive marks (left) and the corresponding probability distribution of observing given number of nucleosomes with repressive marks (right). All simulations are performed with $\Delta\epsilon=2$, but different $J_{\alpha\alpha}$ values, (a): $J_{\alpha\alpha}=0$, (b): $J_{\alpha\alpha}=2.5$, C: $J_{\alpha\alpha}=3.5$.  Other parameters values are from Table 1. The dwelling time distribution is obtained by averaging over $100$ trajectories, each started with a randomly selected initial histone modification configuration, simulated for $10^3$ Gillespie steps, then followed by another $2\times 10^3$ Gillespie steps for sampling.
}
\label{fig:traj}
\end{figure}

To understand the molecular mechanism underlying the dynamics, we hypothesize that the enzyme lateral interactions are essential for collective nucleosome modification. Indeed, Fig. \ref{fig:traj} shows that with $J_{\alpha\alpha}=0$  the percentage of nucleosomes with repressive marks fluctuates but shows a unimodal distribution. A cell with this dynamical property cannot maintain a memory of its epigenetic state over generations. With  $J_{\alpha\alpha}=2.5$, however, one can identify clearly a two-state dynamics from the trajectories, which is further evidenced by the bimodal distribution of the fraction of time the system stays at a collective epigenetic state. A cell gaining  certain epigenetic pattern can propagate the information to its progenies for  several generations before losing it. With an even larger $J_{\alpha\alpha}=3.5$, a cell can stay in one epigenetic state over many cell cycles, and the states with high and low fraction of repressive marks are well separated.
In these calculations the parameters for active and repressive modifications are the same, therefore the behavior of active marks is similar but anti-correlated with that of the repressive marks.

Close examination of the trajectory in Fig. \ref{fig:traj}(c) reveals that a major contribution to  nucleosome mark fluctuations is due to  random replacements during  every cell cycle.
After each cell division, the fraction of repressive marks relaxes quickly to a steady state value before the next cell division. Figure S7 shows the relaxation time is about 6 hours, which is also consistent with experimental measurement on HeLa cells \cite{xu2011model}. It is natural to conjecture that this fast relaxation (less than one cell cycle) is necessary for maintaining a stable epigenetic state against cell division perturbation. We also define an average dwelling time at an epigenetic state as the average time the system stays in the epigenetic state with one mark dominating before it switches to the state with another mark dominating; this is calculated using the algorithm adapted from ref.\cite{Canny1986}. Figure S6 shows that it increases with the cell cycle time. That is, shorter cell cycle makes the epigenetic state less stable. This is consistent with experimental findings that increasing cell division rate accelerates the epigenetic reprogramming from differentiated cells to induced pluripotent stem cells \cite{Hanna2009}.

\begin{figure}
\centering\includegraphics*[width=3.3in,keepaspectratio=true]{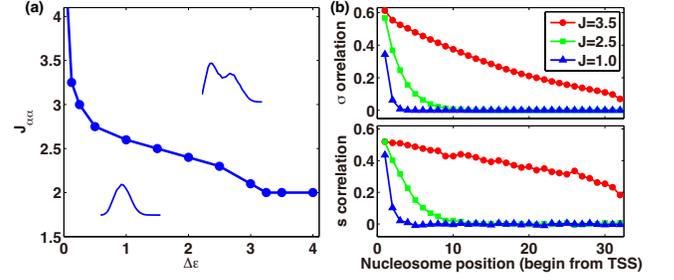}\\  \caption{(Color online) Mechanism of bistability. (a): Phase diagram on the $\Delta\epsilon$-$J$ plane. All other parameters take values in Table 1. (b): Correlation functions $\bar{C}_{1,1}(\sigma_0,\sigma_{L})$ (upper) and $C_{1,1}(s_0,s_{L})$ (lower). }
\label{fig:phase}
\end{figure}

To further analyze the dependence of the model bistable behavior on parameters, we explore the bistable region in the $\Delta\epsilon$-$J$ plane. Figure \ref{fig:phase}(a) shows that a finite value of $J$ is necessary for generating bimodal distributions of the fraction of histones with repressive marks. Below a critical value $\sim 2$, the system only shows unimodal distribution even with very large $\Delta\epsilon$ values.  The required value of $J$ also increases sharply upon decreasing $\Delta\epsilon$. With $\Delta\epsilon=0$, the system can not generate a bimodal distribution with an arbitrarily large value of $J$. While one should be cautious of results with large (possibly unphysical) values of $J$ and $\Delta\epsilon$ since the time-scale separation argument then becomes questionable, the results in Fig. \ref{fig:phase}(a) suggest that both $J$ and $\Delta\epsilon$ are necessary to generate bimodal distributions.

The above results demonstrate that the model, which is based on only nearest-neighbor enzyme lateral interactions without direct correlation of $s$-state update dynamics between two nucleosomes, can generate the observed inheritable epigenetic bistability. Does it contradict with the nonlocal interaction requirement of previous studies  \cite{Dodd2007,Sedighi2007, binder2013transcriptional, Angel2011}? To gain mechanistic understanding, we present a statistical analysis on possible correlations between different nucleosomes. First we define the correlation function for the $\sigma$ states of two nucleosomes $i$ and $j$ with a given set of $s$ configurations, \textit{i.e.}, the correlation between nucleosome $i$ in state $\sigma_i = \alpha$ and nucleosome $j$ in $\sigma_j = \beta$,
\begin{eqnarray}
&&C_{\alpha,\beta}(\sigma_i,\sigma_j; \{s\})
=\frac{ \langle \delta_{\sigma_i, \alpha}\delta_{\sigma_j,\beta} \rangle_{s} -  \langle \delta_{\sigma_i, \alpha}\rangle_{s}  \langle  \delta_{\sigma_j,\beta} \rangle_{s}}
                { \langle \delta_{\sigma_i, \alpha}\delta_{\sigma_i,\alpha} \rangle_{s} -  \langle \delta_{\sigma_i, \alpha}\rangle_{s} \langle  \delta_{\sigma_i,\alpha} \rangle_{s}} \nonumber\\
              &&  =  \frac  {P_{\sigma_i = \alpha, \sigma_j=\beta}(\{s\}) - P_{\sigma_i = \alpha}(\{s\})  P_{ \sigma_j=\beta}(\{s\})  }  {P_{\sigma_i = \alpha}(\{s\})(1 - P_{\sigma_i = \alpha}(\{s\}))   }. \nonumber
\end{eqnarray}
For $j-i=1$, \\{\small$P_{\sigma_i,\sigma_j}(\{s\}) =Tr[ \mathbf{T}_1\cdots \mathbf{T}_{i-2}\mathbf{G}^3_{\sigma_i=\alpha,\sigma_{i+1}=\beta} \mathbf{T}_{i+2}\cdots\mathbf{T}_{N}]/Z_s$,}
and for $j-i>1$,\\
{\small$P_{\sigma_i,\sigma_j} (\{s\})=Tr[ \mathbf{T}_1\cdots \mathbf{T}_{i-2}\mathbf{G}_{\sigma_i} \mathbf{T}_{i+1}\cdots \mathbf{G}_{\sigma_j} \mathbf{T}_{j+1}\cdots\mathbf{T}_{N}]/Z_s$,}
where $\langle \cdot\rangle_{\{s\}}$ means ensemble average with a given set of $s$ configurations, $\delta$ is the Kronecker delta function, $\mathbf{G}^3_{\sigma_i=\alpha,\sigma_{i+1}=\beta} $ is obtained by replacing all the elements in $\mathbf{T}_{i-1} \mathbf{T}_i \mathbf{T}_{i+1}$ containing no term related to $\sigma_i, \sigma_{i+1}$ by zero, {\textit i.e.}, only keeping the $(\sigma_i)$-th column of $\mathbf{T}_{i-1} $,
the $(\sigma_i)$-th row and  $(\sigma_j)$-th column of  $\mathbf{T}_{i} $
and  $(\sigma_j)$-th row of $\mathbf{T}_{i+1}$ nonzero.

Averaging over $N_s$ consecutive samples of Gillespie simulations with the waiting time at each step (the time it takes for the next Gillespie move) $\tau_l$, and the total simulation time $t=\sum_{l=1}^{N_s}\tau_l$, we obtain the correlation functions averaged over the $s$ states,
\begin{eqnarray}
\bar{C}_{\alpha,\beta}(\sigma_i,\sigma_j) &=& \sum_{l=1}^{N_s}  C_{\alpha,\beta}(\sigma_i,\sigma_j; \{s\}) \tau_l/t.
\end{eqnarray}

Similarly we define the $s$ state correlation functions as
\begin{eqnarray}
C_{a,b}(s_i,s_j) &=&\frac{ \langle \delta_{s_i, a}\delta_{s_j,b} \rangle -  \langle \delta_{s_i, a}\rangle\langle  \delta_{s_j,b} \rangle} { \langle \delta_{s_i, a}\delta_{s_i,a} \rangle -  \langle \delta_{s_i, a}\rangle\langle  \delta_{s_i,a} \rangle},
\end{eqnarray}
where with $N_s$ samples, $ \langle \delta_{s_i, a}\rangle = \sum_{l=1}^{N_s} \delta_{s_i(l), a} \tau_l/t$,
with corresponding definitions for other terms.

The nucleosome enzyme binding states show correlations from the smallest length scale, nearest neighbors for small \textit{J} values, to the larger length scales spanning the whole region for sufficiently large \textit{J} values (Fig. \ref{fig:phase}(b)). It is not surprising for a Potts-type model with nearest-neighbor interactions to give rise to beyond-nearest-neighbor correlations of $\sigma$ states. Because the $\sigma$ and $s$ state dynamics are coupled, the $s$ states of nucleosomes also show similar correlations. This nonlocal nucleosome-nucleosome $s$ state correlations are mediated through enzyme binding.

In summary, our model analysis shows that the experimentally observed nearest-neighbor interaction and modification state biased enzyme recruitment of individual nucleosomes work synergetically and sufficiently to result in collective  active and repressive epigenetic states. Unlike a simple 1-D model with nearest neighbor interactions that shows no phase transition, the coupled two-layer model here gives rise to bistability due to positive feedback of nucleosome mark state to enzyme recruitment.
The model supports the proposal of Dodd \textit{et al.} \cite{Dodd2007} that nonlocal 'effective interactions' among nucleosomes affect the covalent modification rates (as evidenced by the dependence of $P_{\sigma_i}$ on $s$ states of all nucleosomes) and are necessary for generating robust bistable epigenetic states. In the supporting text we compare the two models. Our analysis demonstrates a possible molecular mechanism of generating these effective interactions, and epigenetic memory, mediated through nearest-neighbor enzyme lateral interactions. Let's focus on a specific unmarked nucleosome. Without interactions from other nucleosomes, with a set of symmetrically chosen parameters the nucleosome has equal probability of being actively or repressively modified. The term $\Delta\epsilon$ determines what types of enzymes are likely to bind on other nucleosomes within the correlation region. The enzyme lateral interactions ($J$) result in the stabilization of enzyme binding on this tagged site by the binding events at other nucleosomes within the correlation region. This allows the nucleosome to ``read" the majority epigenetic mark type of these nucleosomes, bias its recruitment of  the corresponding enzyme and ``write" on itself accordingly. As shown in the online supporting text, this mechanism is robust with different choices of  model parameter values, with the essential requirement that the time scale for mark restoration must be faster than that of perturbations, mainly from mark removal reactions, cell division and histone exchange, all of which may vary significantly among different cell types.

Our analysis does not rule out other possible mechanisms for epigenetic memory, such as direct interactions among distant nucleosomes due to compact histone structure \cite{Ruthenburg2007}. Inclusion of these interactions extends the present one dimensional two-layer Potts model into higher dimensions and one expects even richer physics \cite{Yujin}. Furthermore, epigenetic memory is maintained by a closed network coupling regulations at different levels including gene expression, epigenetic modification, chromatin remodeling, \textit{etc} \cite{Padinhateeri2011, Ptashne2013}, and requires an integrated treatment in the future.

We thank Drs Andrew Angel, Bing Zhu and Michael Surh for their careful review of our manuscript and insightful comments. We also thank Mr. Yujin Kim for
discussions. This work has been supported by National Science Foundation Grants DMS-0969417,  EF-1038636, and DGE-0966125.


\end{document}